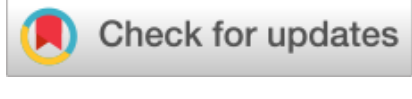

RESEARCH ARTICLE

# Forecasting financial markets with semantic network analysis in the COVID-19 crisis

**Andrea Fronzetti Colladon**[1] | **Stefano Grassi**[2] | **Francesco Ravazzolo**[3] | **Francesco Violante**[4]

[1]Department of Engineering, University of Perugia, VIa G. Duranti 93, Perugia, Umbria, 06125, Italy

[2]University of Rome 'Tor Vergata', Via Cracovia 50, Rome, 00133, Italy

[3]Department of Data Science and Analytics, BI Norwegian Business School and Free University of Bozen-Bolzano, Nydalsveien 37, Oslo, 0484, Norway

[4]Laboratoire de Finance et assurance, ENSAE—École nationale de la Statistique et de l'Administration Économique - Institut Polytechnique de Paris, CREST, Institut Louis Bachelier and CREATES, 5 Avenue Henry Le Chatelier, Palaiseau, cedex, 91764, France

**Correspondence**

Francesco Ravazzolo, BI Norwegian Business School, Free University of Bozen-Bolzano and RCEA, Bolzano, Italy.
Email: francesco.ravazzolo@unibz.it

**Funding information**

University of Perugia, Grant/Award Number: RICBA19LT; University of Rome 'Tor Vergata', Grant/Award Number: E84I20000900005; Italian Ministry MIUR, Grant/Award Number: 2017TA7TYC; French National Research Agency (ANR); Investissements d'Avenir, Grant/Award Number: LabEx Ecodec/ANR-11-LABX-0047; Danish Council for Independent Research, Grant/Award Number: 1329-00011A

**Abstract**

This paper uses a new textual data index for predicting stock market data. The index is applied to a large set of news to evaluate the importance of one or more general economic-related keywords appearing in the text. The index assesses the importance of the economic-related keywords, based on their frequency of use and semantic network position. We apply it to the Italian press and construct indices to predict Italian stock and bond market returns and volatilities in a recent sample period, including the COVID-19 crisis. The evidence shows that the index captures the different phases of financial time series well. Moreover, results indicate strong evidence of predictability for bond market data, both returns and volatilities, short and long maturities, and stock market volatility.

**KEYWORDS**

bond and stock markets, forecasting, semantic network analysis

## 1 | INTRODUCTION

In order to make informed decisions, individuals attempt to anticipate future movements of economic variables and business cycle dynamics. This is particularly important for investment decisions in financial markets. An agent can gain from predicting future market needs and investing early, satisfying market demands when they appear. There is a large body of literature on predicting financial markets and many indicators have been used as predictors. However, the predictability power of such indicators is often questionable and evidence of systematic predictability is weak; see Welch and Goyal (2008).

More recently, a new theory based on a different type of data has received increasing attention, namely, the news media perception and evaluation of the business cycles; see Beaudry and Portier (2006). Beaudry and Portier (2014) state that "according to the news media view of the business cycle, both the boom and the bust are direct consequences of people's incentive to speculate on information related to future developments of the economy." One of the main challenges of this theory is the definition of news and how to empirically measure it. To overcome the problem, Larsen and Thorsrud (2019) propose a novel and direct measure of media news based on their semantic content. Using text data from a Norwegian financial newspaper, they document a superior predictability power of their indicator for Norwegian stock indices returns.

The recent COVID-19 crisis has rendered the task of predicting financial market fluctuations even more difficult. Markets have promptly reacted to early news related to the coronavirus. Indeed, as early as March 2020, markets collapsed, with weekly losses above 30%, well before macroeconomic indicators were impacted, as well as accurate information about the severity of the spread of the coronavirus in Europe and in the United States was available. In this context, Italy represents a peculiar case. Indeed, it was the first country in Europe to experience a major outbreak of COVID-19, with a much higher mortality rate than observed in other countries. Such period of economic and social turmoil, where the news media have not merely covered the role of broadcasting information but also that of conveying perceptions and expectations about future states of the economy, represents an unprecedented testing ground to evaluate the link between news media information and macrofinance variables.

This paper introduces a new index of semantic importance. The index is based on a novel methodology that evaluates the relative importance of one or more general economic-related keywords (ERKs) that appear in the news. The index, whose construction combines methods drawn from both network analysis and text mining, evaluates semantic importance along the three dimensions of prevalence, that is, frequency of word occurrences; connectivity, that is, degree of centrality of a word in the discourse; and diversity, that is, richness and distinctiveness of textual associations. Previous research mainly looked at media sentiment or media coverage, without analyzing the embeddedness that ERKs have in the corpus and their relationships with other words. Fronzetti Colladon et al. (2020) suggest that sentiment can be much less informative than semantic importance. This is sometimes attributable to the methodologies used to calculate sentiment, for example, when trained on a general domain and then applied to a specific context. Indeed, sentiment algorithms have variable error rates, and their reliability has been questioned in different studies (Jussila et al., 2017). Accordingly, our approach excludes sentiment from the main indicator, thus measuring importance, and not favorability, of ERKs.

This approach is also new in the fact that it does not only primarily consider the frequency of ERKs (e.g., Akita et al., 2016; Sun et al., 2016) but it adds information from their relationships with the other concepts in the text. Indeed, we work on co-occurrence networks and apply methods of social network analysis for text mining. We follow a conceptualization where the importance of a term is determined by its frequency, the number and uniqueness of its associations, and its "brokerage" power (Fronzetti Colladon, 2018). To the extent of our knowledge, this approach has never been used in the context of financial forecasting. It also differs from other approaches that use word embeddings (e.g., Oncharoen & Vateekul, 2018; Vargas et al., 2017) and other methods requiring a training set for supervised machine learning, which are sometimes more challenging to interpret.

We identify 38 relevant ERKs suited to our scope. Using a large database of articles published by a pool of Italian newspapers, over the period spanning between January 2017 and August 2020, we assign a score to each ERK, compounding the three dimensions mentioned above. We then aggregate the information from the 38 ERKs in a single composite news index. For the aggregation, we apply partial least squares (PLS) between the target variable and the (38 ERKs) predictors, incorporating information from both the definition of scores and loadings. de Jong (1993) shows that the scores and loadings can be chosen so as to maximize the covariance between the dependent variable and the predictors. To do so, our methodology based on PLS allows us to construct a composite index in a target specific manner. We also consider a selection of the proxy using automatic proxy selection and disciplining variables (factor proxies) based on statistical comparison argument by applying the three-pass filter of Kelly and Pruitt (2015).

While existing empirical literature in this context has focused on stock market return predictability, see Baker and Wurgler (2006), Baker and Wurgler (2007), Chung et al. (2012), and Limongi Concetto and Ravazzolo (2019) among others, we evaluate the power of our media news index in predicting not only the Italian stock market aggregate return but also various short and long maturity government bonds index returns, as well as their volatility.

Periods of large movements in the stock and bond markets are associated with political instability and economic uncertainty. Our findings show that the index is able to anticipate the different phases of the market and capture idiosyncratic features of each series. We find evidence of economically meaningful and statistically significant predictability for government bond returns and volatilities. For stock market data, we find evidence of predictability of the market portfolio returns only to a limited extent. However, when predicting stock market volatility, adding information contained in the news media improves the prediction accuracy up to 9%, compared with standard benchmark forecast models. Alternative (standard) methods for dealing with newspaper information, such as the sentiment index, do not offer similar gains.

The remainder of the paper is organized as follows. Section 2 introduces our new textual data index. Section 3 provides a detailed description of the data employed, the methodological strategy used to predict financial time series with textual data, and the results of our analysis. Section 4 complements them with an economic evaluation of the gains. Finally, Section 5 concludes. Appendix A1 reports further results.

## 2 | A NEW INDEX FOR TEXTUAL DATA

In this paper, we propose a novel measure of semantic importance that combines methods drawn from both social network analysis and text mining. The index aims at measuring the relative importance of a predefined set of words mentioned in a large set of textual documents. The methodology labeled Semantic Brand Score was introduced by Fronzetti Colladon (2018) for application to commercial brands' reputation and awareness but has never been applied in the economic and financial environment. Starting from the word frequency as a natural measure of importance within a text (Piantadosi, 2014), the associations that a word has in the text, as well as the heterogeneity of its context, are used as pivotal additional variables for a comprehensive assessment. Our index explicitly exploits the relationships among words in a text. To this end, texts are transformed into networks of co-occurring words and relationships are studied through social network analysis, see Wasserman and Faust (1994). As an example, consider the following sentence (and a word co-occurrence threshold of three words) to generate the network reported in Figure 1: "The proud and unfeeling landlord views his extensive fields, and without a thought for the wants of his brethren, in imagination consumes himself the whole harvest" (from The Theory of Moral Sentiments of Adam Smith). Words are presented without stemming, for the sake of readability.

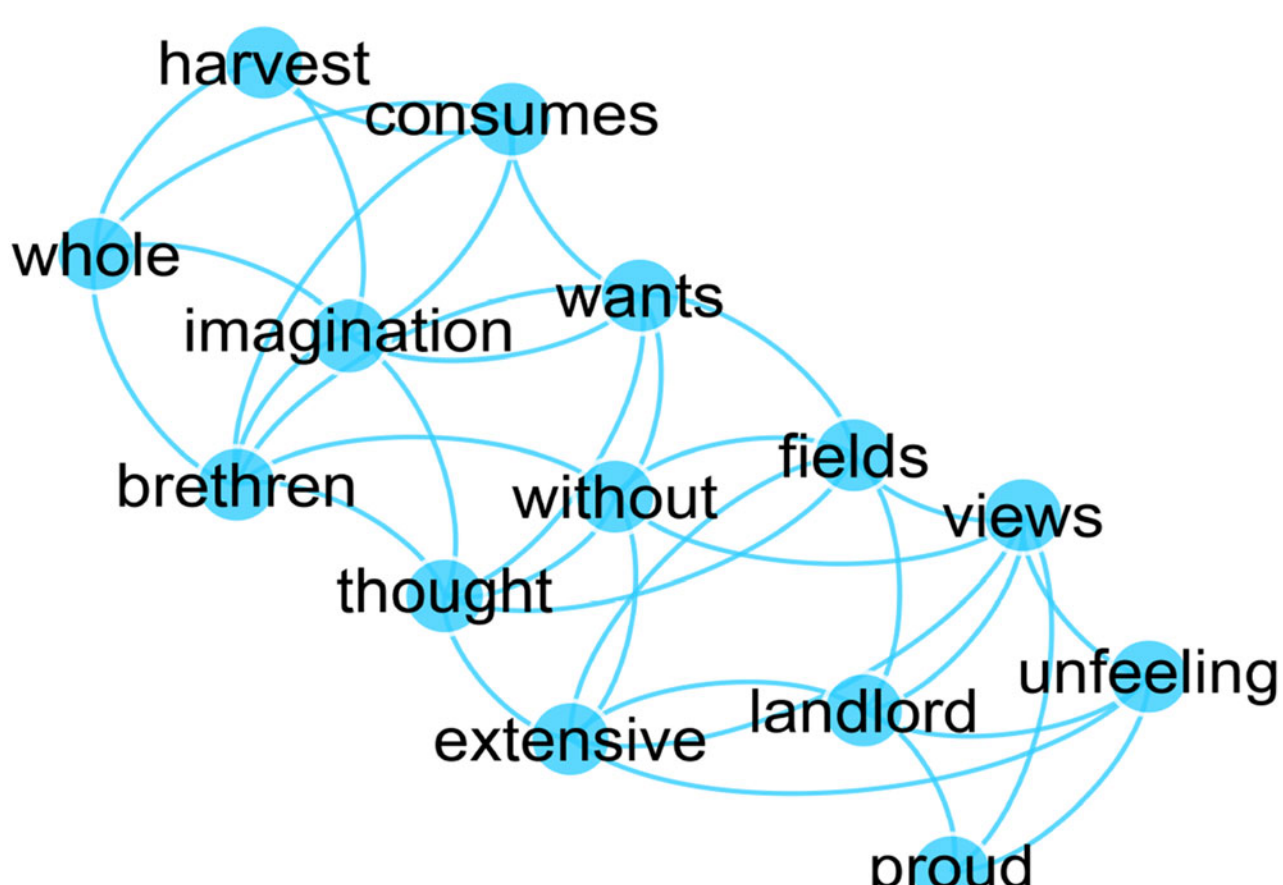

FIGURE 1 Graphical representation of the sentence "The proud and unfeeling landlord views his extensive fields, and without a thought for the wants of his brethren, in imagination consumes himself the whole harvest" (from The Theory of Moral Sentiments of Adam Smith)

The index measures words' semantic importance, that is, in our context of the selected ERKs, along the three dimensions of prevalence, diversity, and connectivity.

Prevalence, which relates to the notion of awareness, see Keller (1993), measures how frequently an ERK is mentioned, the rationality being that an ERK used frequently is easier to remember and more recognizable. Prevalence is calculated as the frequency of a word in a given set of documents and time frame. Prevalence of a particular set of words could ultimately influence the opinions and behaviors of the readers. For instance, in our context, the recurrent use of specific combinations of words may trigger fear or an optimistic view about the current, as well as future, states of the economy.

Diversity measures the degree of heterogeneity of the semantic context in which a word is used, with emphasis on the richness and distinctiveness of its textual associations. Diversity is defined by the number and uniqueness of connections a word has in the co-occurrence network and it is measured by the distinctiveness centrality metric introduced in Fronzetti Colladon and Naldi (2020). More precisely, in a graph of $n$ nodes (words) and $E$ edges (e.g., 1), distinctiveness of node $i$ is given by

$$D_i = \sum\nolimits_{j=1, j \neq i}^{n} \log \frac{(n-1)}{g_j} \mathrm{I}(w_{ij} > 0), \quad (1)$$

where $W$ is the set of weights associated to each edge; $g_j$ is the degree of node $j$, which is a neighbor of node $i$; and $\mathrm{I}(\cdot)$ is an indicator function that equals 1 if there is an edge that connects nodes $i$ and $j$ with positive weight, $w_{ij}$.

We postulate that the degree of diversity provides relevant information about how pervasive a topic is in the weave of the economy and could ultimately attract the attention of a variety of economic actors, for example, institutional investors, policy makers, and private investors.

The third dimension of the index, that is, connectivity, assesses the weighted betweenness centrality of the ERKs; see Brandes (2001) and Freeman (1978). Connectivity measures how much a word is embedded in a discourse acting as a bridge between its parts, or more specifically, how often a word appears in-between the network paths which interconnect the other words in the text. Following Wasserman and Faust (1994), for node $i$, we have:

$$C_i = \sum_{j<k} \frac{d_{jk}(i)}{d_{jk}}, \quad (2)$$

where $d_{jk}(i)/d_{jk}$ is the proportion of shortest network paths connecting nodes $j$ and $k$ (measured by edge weights) that include the node $i$. Finally, an index is constructed as a composite score obtained by summing the standardized measures of prevalence, diversity, and connectivity discussed above. The standardization is carried out considering the semantic network of each time period.

## 3 | FORECASTING STOCK AND BOND MARKETS

The objective of our paper is to assess whether, and if so to what extent, the information contained in textual news data improves the predictability of macrofinance variables. Focusing on the Norwegian stock market, Larsen and Thorsrud (2019) find strong evidence of asset prices predictability by textual news data. They use a latent Dirichlet allocation model that statistically categorizes the corpus, that is, the whole collection of words and articles, into topics that best reflect the corpus's word dependencies. Building on the aforementioned work, we rely on our more general semantic importance index, described in Section 2, as the aggregate measure of textual news data information content. Our empirical evaluation centers on the Italian stock and bond markets. This choice is not coincidental. Italy is a large economy, that is, the eighth world largest economy by GDP, with a fairly large stock market, among the most liquid in Europe and, with the third largest sovereign debt in the world after Japan and US. This makes the Italian government debt market very attractive, and thus liquid, at all maturities. We assess the predictive power of textual news information as a nontraditional driver of the returns level, as well as volatility, of five stock and bond time series.

Our set of target variables comprises the aggregate market portfolio, that is, FTSE MIB index, and 2-, 5-, 10-, and 30-year maturity Italian government bond indices, collected from Datastream. The bond indices are maintained by Refinitiv and are designed to track the performance of euro-denominated securities publicly issued by Italy for its domestic market. Such indices provide high-quality measurements of bonds with similar maturities available in the market. For each of these variables, available daily, we compute logarithmic returns and realized volatilities (measured in percentage), aggregated at a weekly frequency. We opt for a weekly aggregation frequency following evidence in Fronzetti Colladon (2020), which showed that daily news data is highly variable and proved that the effect produced by multiple news in 1-week has more impact on citizens' behavior. Indeed, investors can take time to form an opinion on newspaper contents and can adjust their allocation only after sometimes. Moreover, online news is often accessed even several days after publication and can therefore extend its influence over time.

Weekly realized volatility is computed using the range estimator of Parkinson (1980), that is, $(4 \log 2)^{-1} \log (H_t/L_t)^2$, where $H_t$ and $L_t$ represent the highest and lowest prices of week $t$, respectively. The sample period spans from January 6, 2017, to August 28, 2020, totaling 191 weekly observations.

Table 1 provides descriptive statistics for our 10 target series also plotted in Figure 2.

Over the period analyzed, stock market returns are negative on average and, as expected, more volatile than bond returns. The lowest stock market return (−37%) occurs in the week between Monday, March 9, 2020, and Friday, March 13, 2020, that is, the inception of the COVID-19 crisis. Bond returns for maturities longer than 2 years are positive on average, likely boosted by the quantitative easing program initiated by the European Central Bank, and exhibit an upward sloped volatility term structure.

**TABLE 1** Descriptive statistics (mean, standard deviation, maximum, and minimum) of the weekly FTSE MIB and the 2-, 5-, 10-, and 30-year maturity BTP indices return and realized variance

| | Returns | | | | Realized variance | | | |
|---|---|---|---|---|---|---|---|---|
| | **Avg** | **SD** | **Max** | **Min** | **Avg** | **SD** | **Max** | **Min** |
| FTSE MIB | −0.004 | 3.795 | 11.600 | −36.961 | 0.101 | 0.378 | 5.080 | 0.001 |
| BTP-2y | −0.009 | 0.301 | 1.829 | −1.240 | 0.001 | 0.001 | 0.127 | 0.000 |
| BTP-5y | 0.027 | 0.834 | 4.123 | −3.967 | 0.008 | 0.031 | 0.348 | 0.000 |
| BTP-10y | 0.060 | 1.327 | 4.389 | −6.855 | 0.020 | 0.050 | 0.479 | 0.000 |
| BTP-30y | 0.099 | 2.295 | 8.967 | −9.853 | 0.053 | 0.106 | 1.117 | 0.002 |

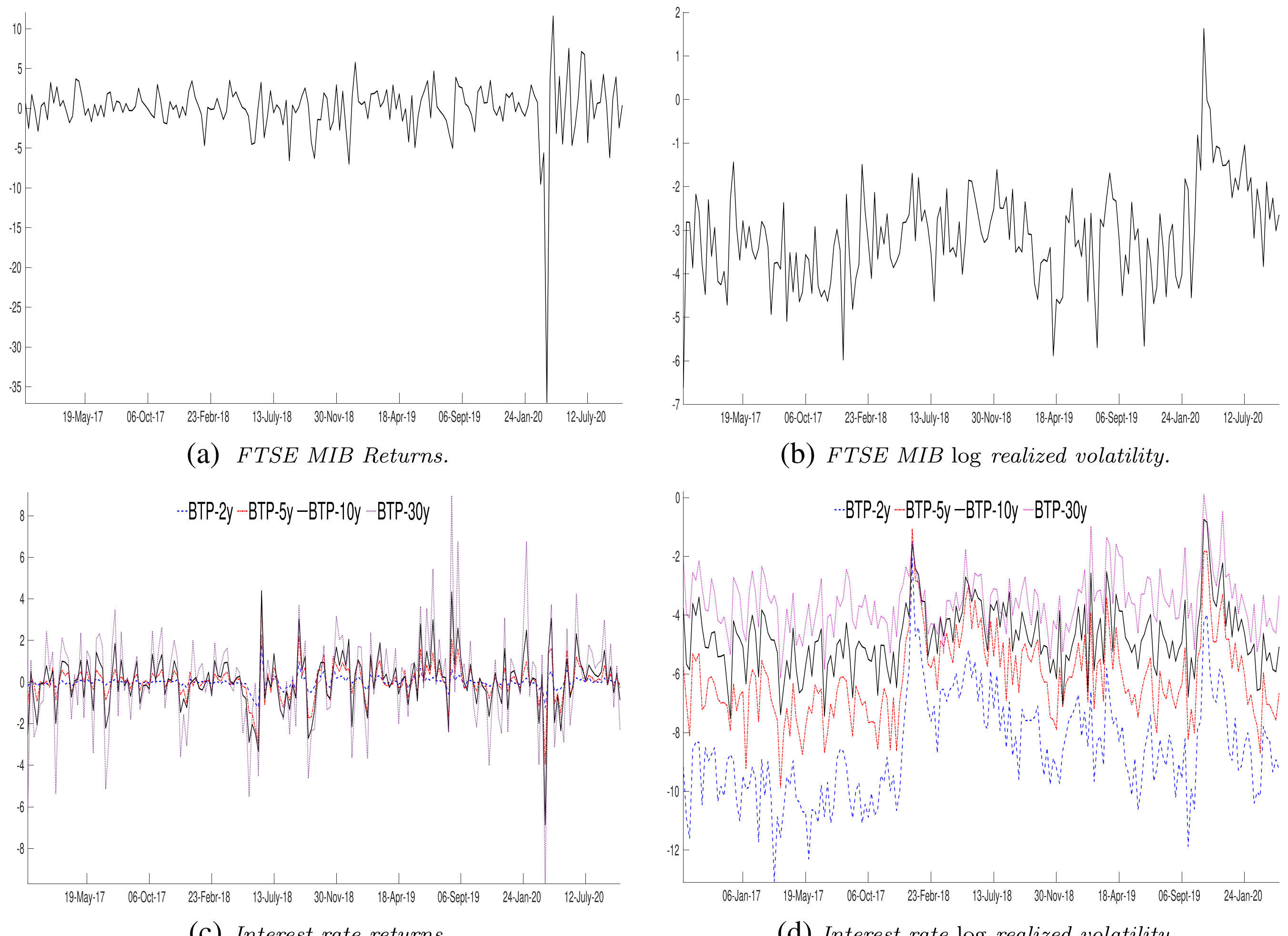

**FIGURE 2** Dataset. The figure reports the dataset used in the forecasting study. Subplot (a) reports the FTSE MIB returns; subplot (b) reports the FTSE MIB volatility; subplot (c) reports the BTP-2y, BTP5y, BTP-10y, and BTP30y returns; subplot (d) reports the BTP-2y, BTP5y, BTP-10y, and BTP30y volatility

## 3.1 | Textual data collection and key words

Choosing pertinent keywords to search in a database of newspapers articles is crucial for the construction of an informative textual index. As documented in literature, such choice is nontrivial because word meaning can vary across fields and users. For instance, Loughran and McDonald (2011) find that words associated to a negative attribute or meaning by widely used dictionaries, for example, the Harvard Dictionary, are words typically not considered negative in financial contexts. To circumvent the problem we select a limited set of words with a clear economic meaning, homogeneously understood by the larger public.

**TABLE 2** List of economic-related keywords

| | | | | | | | |
|---|---|---|---|---|---|---|---|
| Word singletons | | | | | | | |
| 1 | Spread | 7 | Quantitative easing | 13 | Rating | 19 | Real economy |
| 2 | Interest rates | 8 | Monetary policy | 14 | Eurogroup | 20 | European Commission |
| 3 | Euro | 9 | Bank of Italy | 15 | coronabond | 21 | Eurobond |
| 4 | European troika | 10 | ESM[a] | 16 | SURE[b] | 22 | EIB[c] |
| 5 | Junk bond | 11 | Oil | 17 | Gold | 23 | Financial markets |
| 6 | Strikes | 12 | INPS[d] | 18 | GDP | 24 | Confindustria[e] |
| Word sets | | | | | | | |
| 1 | COVID, coronavirus | 5 | BTP, BOT, CCT[f] | 9 | savings, savers | 13 | european union, EU |
| 2 | lockdown, quarantine | 6 | inflation, prices | 10 | deficit, gov.t debt | 14 | consumption, -umers |
| 3 | taxes, taxation, wealth tax | 7 | Borsa Italiana, FTSE MIB, FTSE MIB | 11 | unions, CISL, CGIL, UIL[g] | | |
| 4 | economic crisis, recession, economic pandemic | 8 | unemployment, redundancy, unemployment benefit | 12 | smart working, distance work | | |

[a]European stability mechanism.
[b]European instrument for temporary support to mitigate unemployment risks in an emergency.
[c]European Investment Bank.
[d]Italian social welfare and pension institution.
[e]Italian industrial sector association.
[f]Italian debt instruments.
[g]Acronyms of the three largest Italian trade unions.

We choose 38 sets of keywords: 24 singletons, that is, individual words, albeit included also in plural form when it exists, and 14 sets of words sharing similar meaning, that is, synonyms, or identifying similar items or meaning. The set of keywords is reported in Table 2, translated from Italian where needed.

The textual data used in the analysis is provided by Telpress International, and it is collected from multiple online news sources. To generate networks from texts and to calculate our semantic importance index, we rely on the SBS BI web application,[1] see Fronzetti Colladon and Grippa (2020), and the computing resources of the ENEA/CRESCO infrastructure (Ponti et al., 2014). Prior to the computation of the semantic importance index, common text preprocessing routines (Perkins, 2014), such as tokenization, removal of stop words, and removal of word affixes, known as stemming (Willett, 2006), are implemented. Then, a social network based on word co-occurrences is generated for each time interval considered in the analysis.

The database of Italian news, published between January 2, 2017, and August 30, 2020, contains more than 772,500 news articles. For each news, we only consider the title and the lead, that is, the initial 30% of text, ignoring the remaining part.[2] This is consistent with previous work, which suggested that semantic importance indices are more informative when calculated on the news parts that better capture the readers' attention, that is, the title and the lead; see Fronzetti Colladon (2020). This is also aligned with past research, which has already proven that a large part of internet users only read the beginning of online articles, see Nielsen and Loranger (2006), among others.

We calculate an index, as detailed in Section 2, for each of the ERKs listed in Table 2, thus obtaining 38 time series. To reduce uncertainty and aggregate information, following Fronzetti Colladon et al. (2019), we apply PLS between the target variable and the (38 ERKs) predictors,

[1]Available at https://bi.semanticbrandscore.com.

[2]We have also tried using the full content of news, but accuracy was not superior, computational time was substantially longer, and results are not reported.

(a) *Media news index targeted to the returns series*

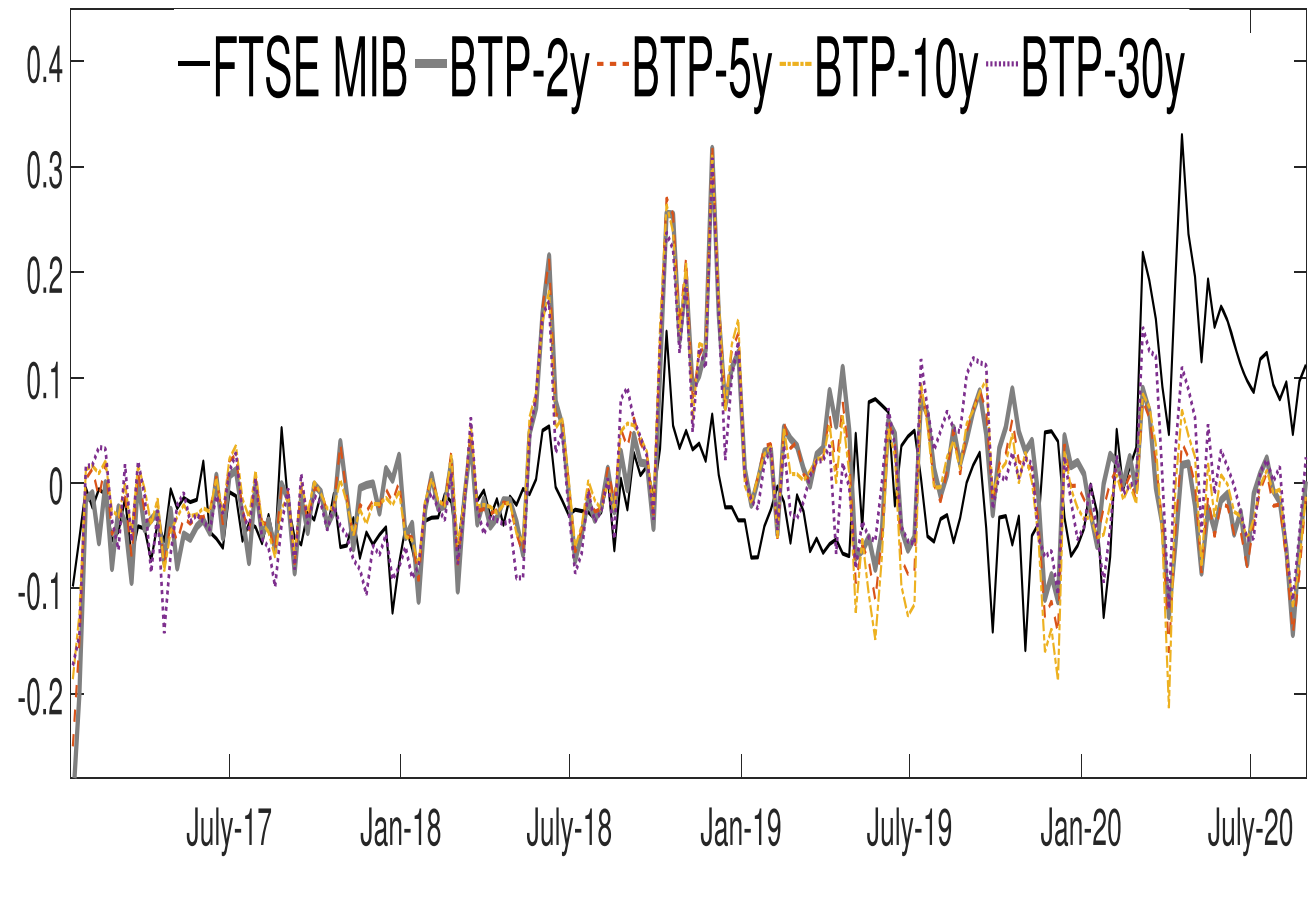

(b) *Media news index targeted to the volatility series*

**FIGURE 3** (a) The semantic importance indices applied to Italian stock returns (FTSE MIB) and Italian bond returns (BTP-2y, BTP-5y, BTP-10y, and BTP-30y). (b) The indices applied to volatility of Italian stock markets and Italian bond markets incorporating information from both the definition of scores and loadings. Therefore, our measure is a series specific index.

Figure 3 shows our full sample indices associated to returns and volatilities target variables. For the sake of comparability, each index is centered and standardized. Figure 3a indicates that the indices targeted on the return series follow similar patterns, somewhat less evident when compared with those targeted to volatility, with large movements associated to destabilizing political and economic events. For example, the sensible negative movement in the index during spring 2018 coincides with elections where no party achieved a sufficient majority; the sharp positive bounce during autumn 2018, instead, reflects the reaction to pension and social welfare reforms, introduced by the government composed by the conservative LEGA and populist M5S parties, that markets and the European Union did not fully support. Similar patterns are observed during the late spring and though summer 2019, period in which the Italian government suffered strong internal disagreements on several affairs, including immigration, and which resolved in a new government coalition. Finally, the effect of the inception of the COVID-19 pandemic is clear.

Figure 3b shows that indices targeted to volatility are more variable than those constructed for returns. Indeed, the average correlation of the latter is 97%, and the average correlation of volatility targeted indices is 47%. However, the occurrence of large shocks and instability is aligned between the two sets of indices. The only striking exception is observed at the end of 2019, when patterns of the two sets of indices series appears to diverge.

## 3.2 | Design of the forecasting exercise

Our forecasting exercise aims at assessing whether including information stemming from news contributes to improving stocks and bonds returns predictions. We employ a recursive forecasting scheme, using an expanding estimation sample, to produce 1-week-ahead forecasts.[3] The first estimation sample spans from January 6, 2017, to April 26, 2019. The out-of-sample (OOS) forecast evaluation period spans the following 70 weeks, that is, from May 3, 2019, to August 28, 2020. Therefore, we consider both the pre-COVID-19 period, the turbulent COVID-19 outbreak period in March and April 2020, and the following less volatility COVID-19 period up to the end of summer 2020.

We opt for a simple forecasting model, that is, the ARX(1):

$$y_{t+1} = \alpha + \gamma y_t + \beta x_t + \varepsilon_{t+1}, \tag{3}$$

where $y_{t+1}$ is the target variable we aim at predicting, $x_t$ is a set of news information predictors, and $\varepsilon_{t+1} \sim WN(0,\sigma^2)$. An obvious choice for $x_t$ is the pool (or a subset) of the 38 index assigned to the ERKs listed in Table 2. If, on the one hand, this approach enables identifying and isolating ERKs with predictive power from the rest, on the other hand, the large set of regressors, relative to the limited number of observations in our sample, may generate an undesirable level of uncertainty.

[3]We present the models and results for one-step-ahead horizon, that is, $h = 1$. The model can be generalized to multistep ahead horizons, $h > 1$. Tables A.1 and A.2 report 2-, 3-, and 4-week-step-ahead results when applying direct forecasting.

To circumvent this problem, we convey the information contained in the index associated with the individual (sets of) keywords using two alternative aggregation approaches. The first approach extracts, from the pool of 38 ERKs variables, one or more common factors by means of PLS.[4] The PLS is computed individually for each series and repeated for each vintage that forecasts are produced, therefore using real-time information. We labeled it as "ERK model" in the remainder of the paper.

Moreover, we explore as an alternative scheme to handle the many ERKs, the three-pass regression filter (3PRF) of Kelly and Pruitt (2015). The 3PRF imposes an intuitive constraint which ensures that the factors irrelevant to **y** drop out of the 3PRF forecast. The formulation of the filter and its success in forecasting relies on the existence of proxies that depend only on target-relevant factors. As describe in Kelly and Pruitt (2015) we consider the automatic proxy-selection algorithm (3PRF-A) and the theory-motivated proxies based on factor proxy (3PRF-T). In this case the model is

$$y_{t+1} = \alpha + \mathrm{F}_t \boldsymbol{\beta} + \varepsilon_{t+1}, \tag{4}$$

where $\mathrm{F}_t$ is the predictive factor based on the 3PRF-A or the 3PRF-T approach.[5]

The competing aggregation method, following Stock and Watson (1999) and Timmermann (2006), consists in computing multiple forecasts for each target variable using Equation (3) and one ERK predictor, $x_{i,t}$, at the time and then in aggregating those forecasts. We use the equal weight combination:

$$y_{t+1} = \sum\nolimits_{i=1}^{38} w_i \hat{y}_{i,t+1}, \tag{5}$$

where $\hat{y}_{i,t+1}$ is the forecast of $y_{t+1}$ generated by the linear ARX(1) using $x_{i,t}$ and $w_i = 1/38, i = 1,...,38$.

We contrast the predictive performance of the model exploiting textual news information against a set of standard benchmarks: The white noise model when the target is the return of either stocks and bonds and the heterogeneous autoregressive model (HAR) model, adapted to weekly series, of Corsi (2009) using the following specification:

$$RV_{t+1,w}^{(w)} = c + \beta^{(w)} RV_t^{(w)} + \beta^{(m)} RV_t^{(m)} + \omega_{t+1,w},$$

where $RV_t^{(w)}$ is the weekly $t$ realized volatility and the monthly aggregate is given by

$$RV_t^{(m)} = \frac{1}{4}\left(RV_t^{(w)} + RV_{t-1}^{(w)} + \ldots + RV_{t-3}^{(w)}\right),$$

when we aim at forecasting volatility.[6] We also consider the HAR with the exogenous variables (HAR-X) where the exogenous variables is the common factor extracted by the PLS discusses above. Furthermore, for both series we apply the autoregressive model of order 1, labeled AR.

Welch and Goyal (2008) document how difficult it is to provide a superior forecast performance than the white noise model when predicting stock returns. De Pooter et al. (2010) show similar evidence for bond markets. This is because prices or yields are nonstationary, or close to, which makes it difficult to outperform the simple no-change forecast. When we turn to volatility instead, the nonstationary component is typically less relevant and the time-reversion is more pronounced, with periods of high and low volatility alternating.[7]

Moreover, we also compare our index to an alternative and well-known text evaluation method: the sentiment index; see Fraiberger et al. (2020). This index is computed by evaluating media sentiment in correspondence with economy-related terms, considering the online news articles included in our study. Prior to the calculation of sentiment, we apply the same text preprocessing procedures described in Section 3.1. Subsequently, we consider the polarity of the words associated to each ERK and compute a weighted average based on co-occurrence strength.[8] Accordingly, the sentiment index may vary from −1 to +1, where negative values represent negative expressions and positive values represent positive expressions. We substitute our textual indicator with the sentiment index in Equation (3) and apply it as an alternative news information predictor. We label it "SI model" in the remainder of the paper.

We measure forecast accuracy by means of mean squared prediction errors (MSPEs). The statistical assessment of the predictive performance differential stemming

[4]We have also tried principal component analysis (PCA), but results were inferior and not reported. We believe that the total number of ERKs is limited and PCA is less adequate in this case.

[5]We have also tried the Lasso approach (Tibshirani, 1996), but results were inferior and not reported.

[6]The white noise model for returns corresponds to the driftless random walk model for price. We refer to it, labeling it RW, in the remaining part of the paper.

[7]We have also applied the RW no-change. Results are substantially inferior to the RW and HAR models and not reported in the text.

[8]We additionally tested other approaches for the calculation of sentiment, such as considering the full articles' content and not just the sentences related to economic terms. However, none of these alternative approaches led to results better than our primary choice.

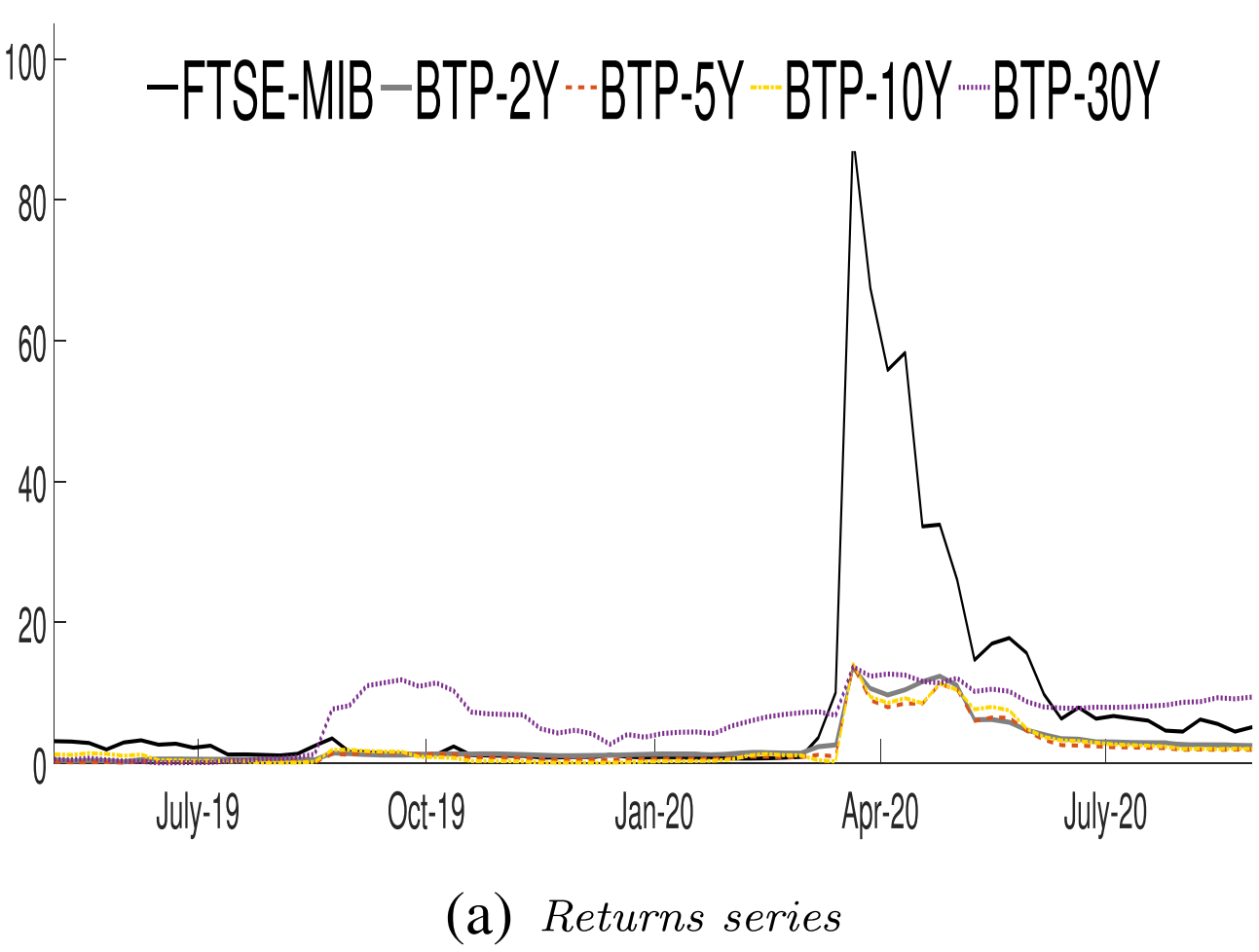

(a) *Returns series*

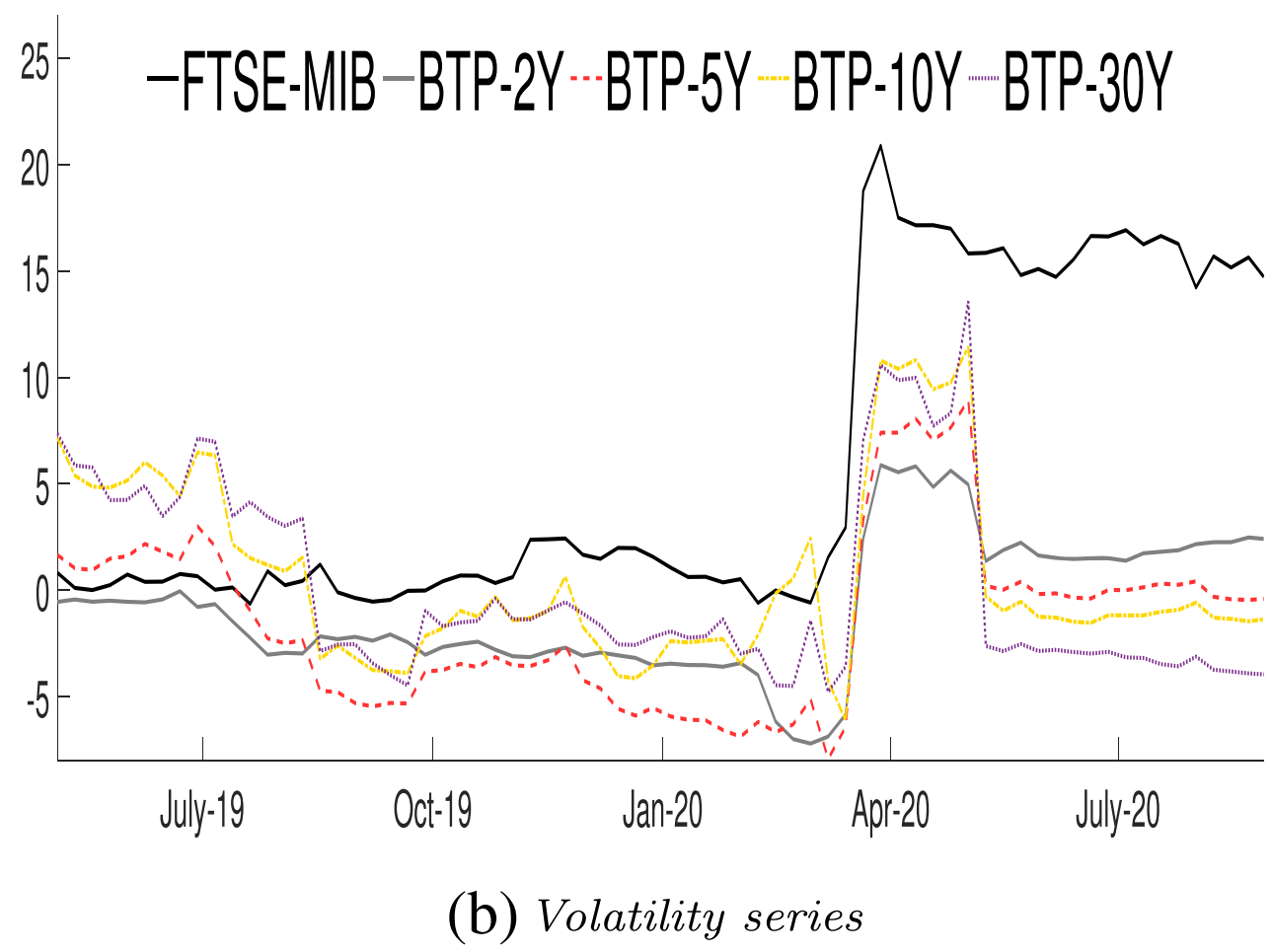

(b) *Volatility series*

**FIGURE 4** (a) Differences in AIC (AIC(AR) − AIC(ERK)) for the AR model without the semantic importance index and the alternative ARX model with the semantic importance index (ERK) when predicting Italian stock returns (FTSE MIB) and Italian bond returns (BTP-2y, BTP-5y, BTP-10y, and BTP-30y); if the AR generates the better fit, then the AIC differences are negative. (b) Differences in AIC (AIC (HAR) − AIC(ERK)) for the HAR model and the alternative ERK when predicting volatility of Italian stock and bond markets from the inclusion of textual news information is based on the Diebold and Mariano test (Diebold & Mariano, 1995, DM).

## 3.3 | Results

### 3.3.1 | In-sample evidence

Results from Inoue and Kilian (2004) imply that in-sample predictability is a necessary condition for OOS predictability. To assess the degree of in-sample fit, in Figure 4, we compare the ERK model based on the semantic importance index to the nested benchmark model by means of Akaike information criterion (AIC). See, for example, Ravazzolo and Rothman (2013) for a similar exercise on the role of oil prices for predictability to US output growth.[9] The AIC is computed recursively for all estimation windows between May 3, 2019 and August 28, 2020.

The inclusion of textual news information improves stock and bond markets predictability in all the sample. The gains are moderate in the first part of the sample. Starting from March 2020, that is, the inception of the COVID-19 crisis, until May 2020, for all the series, the ERK model exhibits a dramatic drop in the AIC, resulting in large positive values for the difference that indicate a large predictability power from our index. The largest difference is observed when predicting stock market returns. Gains continue to exist in the second part of the COVID-19 period after May 2020 associated to lower volatility in the series. The ERK model shows the most persistent gains over the sample for the bond with the longest maturity.

The data shows a similar picture when turning to stock market and bond volatility, albeit less consistent. Until March 2020, for the stock market returns, the ERK is inferior or similar to the benchmark, but then, it dominates for the remaining part of the sample. When turning to bond volatility, in general, the ERK model performs well from summer 2019, a period characterized by political uncertainty in Italy. The ERK performance stabilizes during autumn 2019 after a new government was formed and markets experienced a period of calm until the COVID-19 crisis. The gains are lower in the second part of the COVID-19 period, but the difference remains positive for all four bond comparisons.

*Returns OOS results*

Results in Panel A of Table 3 show that the ERK model that includes the semantic importance of ERKs improves forecasting accuracy for the shorter and longer maturity bond returns. In particular, reduction in MSPE for the 2-year bonds is statistically significant at 10%. MSPE reduction for the 30-year bonds is more moderate and the performance of the RW and ERK models are more similar for the 5-year and 10-year maturity. In the case of

[9]Note that the benchmark model when predicting volatility series is the HAR model. For returns, the benchmark is the RW model. However, because the latter does not have any explanatory component, its AIC is very poor, i.e. it is a function only of the series variance, see the forecasting puzzle in Meese and Rogoff (1983). Therefore, for the sake of a fair comparison, we contrast the AR and ERK models in all ten evaluations.

| Models | FTSE MIB | BTP-2y | BTP-5y | BTP-10y | BTP-30y |
|---|---|---|---|---|---|
| Panel A: Returns | | | | | |
| *RW* | *3.07* | *0.18* | *0.59* | *1.02* | *1.89* |
| AR | 1.07 | 0.99 | 1.00 | 1.01 | 1.00 |
| ERK | 1.02 | 0.98* | 1.02 | 1.01 | 0.99 |
| EW | 1.06 | 0.98* | 1.03 | 1.05 | 1.05 |
| SI | 1.07 | 1.00 | 1.01 | 1.01 | 1.00 |
| 3PFR-A | 1.08 | 1.04 | 1.06 | 1.08 | 1.07 |
| 3PFR-T | 1.18 | 1.07 | 1.09 | 1.12 | 1.05 |
| Panel B: Volatility | | | | | |
| *HAR* | *0.91* | *0.94* | *0.90* | *0.87* | *0.79* |
| AR | 1.08 | 1.00 | 0.99 | 1.00 | 1.03 |
| ERK | 0.92** | 0.98* | 0.97* | 0.98* | 1.03 |
| HAR-X | 0.94** | 1.02 | 1.00 | 0.98* | 1.03 |
| EW | 1.00 | 1.01 | 1.01 | 1.03 | 1.04 |
| SI | 1.01 | 1.02 | 1.02 | 1.02 | 1.03 |
| 3PFR-A | 0.94** | 1.07 | 1.05 | 1.06 | 1.11 |
| 3PFR-T | 0.90** | 0.97* | 1.02 | 1.02 | 1.01 |

**TABLE 3** Panel A provides mean square prediction error (MSPE) results when forecasting Italian stock returns (FTSE MIB) and Italian bond returns (BTP-2y, BTP-5y, BTP-10y, and BTP-30y)

*Note*: Absolute MSPE for the random walk (RW) benchmark is reported (*"in italics"*); relative numbers to the benchmark are given for the alternative autoregressive model (AR), autoregressive extended with the semantic importance index (ERK), equal weight combination of ARX models based on different economic-related keywords (EW), autoregressive extended with the sentiment index (SI), and finally, the three-pass regression filter with automatic proxy selection (3PFR-A) and theory proxy selection (3PFR-T). Panel B provides MSPEs when forecasting the volatility of Italian stock and bond markets. Absolute MSPE for the HAR benchmark is reported (*"in italics"*); relative numbers to the benchmark are given for the alternative specifications: autoregressive model (AR), autoregressive extended with the semantic importance index (ERK), equal weight combination of ARX models based on different economic-related keywords (EW), autoregressive extended with the sentiment index (SI), the HAR model with semantic importance index (HAR-X) and finally the three-pass regression filter with automatic proxy selection (3PFR-A) and theory proxy selection (3PFR-T).

*The alternative model provides superior statistical forecasts at 10% significance level.

**The alternative model provides superior statistical forecasts at 5% significance level.

stock returns (FTSE MIB), results are comparable to the benchmark but not superior. Studying performance over time, the ERK model gives the largest predictability during the first part of the COVID-19 period, but this predictability smooths in the second part of the COVID-19 period for the stock returns and bond returns with middle term maturities, confirming AIC evidence in Figure 4.

The AR model never provides comparable results, and only for the 2-year BTP maturity does it achieve lower MSPE than the RW. The equally weighted forecast combination of ARX (EW) is also statistically superior to the benchmark for the 2-year bond returns, yet falls behind the ERK model. Finally, in none of the five cases does the model based on the sentiment index offer gains similar to the ERK model, and it is never superior to the benchmark RW. Similar evidence is found for the three-pass filter approaches. Therefore, how newspaper information is treated and modeled matters.

### *Volatility OOS results*

Panel B of Table 3 provides results on volatility predictions. In this case, we observe the largest gains, in terms of forecast accuracy, obtained by using the semantic importance index. In four cases over five, the ERK model is statistically superior to the HAR benchmark, with MSPE reductions ranging from 8% to 2%. The large improvement is when forecasting stock market volatility, with a smaller MSPE of 8% and statistically significant at 1% confidence level. When predicting bond volatility, the ERK statistically outperforms the benchmark for 2-, 5-, and 10-year maturities, with the largest economic gains observed for the 5-year maturity (3% improvement). Only for the longest 30 years maturity, there is not evidence of gains. The forecast combination does not measure up, always resulting inferior to the benchmark. In addition, the SI model does not perform similarly to the ERK and it is never superior to the benchmark. This result stresses the importance of the index construction to avoid an unfavorable signal to noise balance. Indeed, textual data help to improve forecast accuracy provided that specific keywords that receive more attention in newspapers are selected. The three-pass filter approaches, 3PFR-A and 3PFR-T, also do well, but in most cases, they are less accurate than ERK and in two cases just marginally more accurate.

Figures 5 and 6 report the OOS performance of models based on newspaper data for the FTSE MIB and BTP10y volatilities, respectively. The figure shows the cumulative squared prediction errors of the prevailing

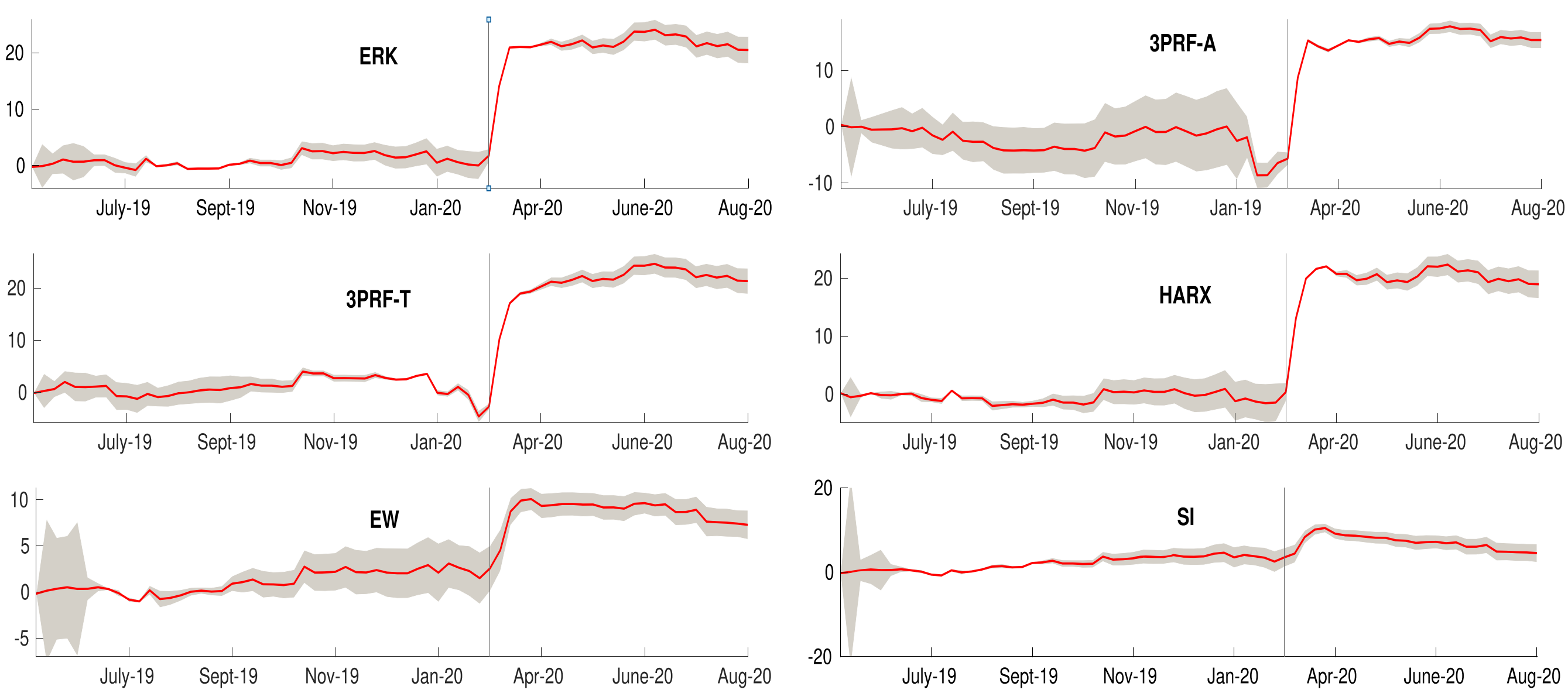

**FIGURE 5** Cumulative squared prediction errors difference for the FTSE MIB volatility. The figure reports the cumulative squared prediction errors of the benchmark model (HAR) minus the cumulative squared prediction error of the alternative models reported in Table 3. The gray band is the equivalent of 95% two-sided levels, based on MSE-T critical values from McCracken (2004)

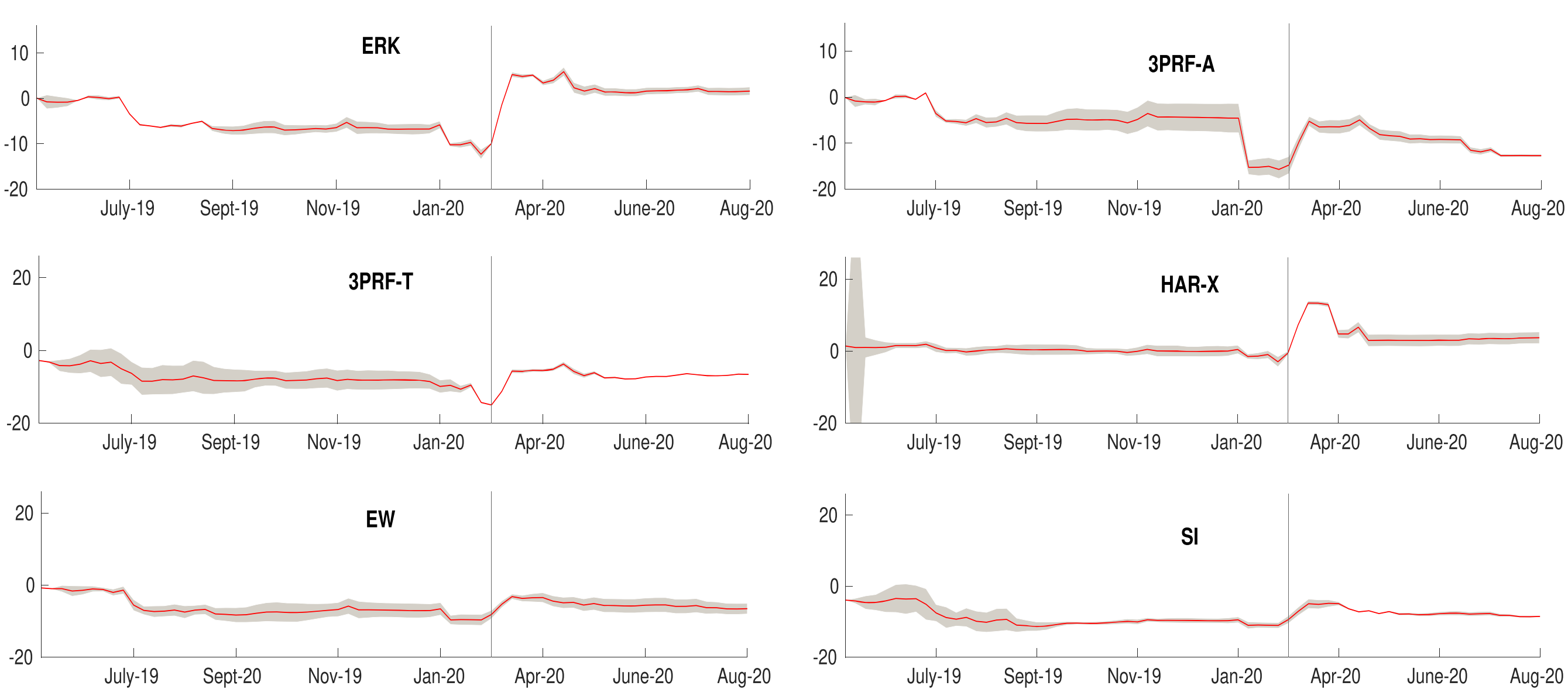

**FIGURE 6** Cumulative squared prediction errors difference for the BTP10y volatility. The figure reports the cumulative squared prediction errors of the benchmark model (HAR) minus the cumulative squared prediction error of the alternative models reported in Table 3; see Welch and Goyal (2008). The gray band is the equivalent of 95% two-sided levels, based on MSE-T critical values from McCracken (2004)

mean minus the cumulative squared prediction error of the alternative models; see Welch and Goyal (2008). Whenever a line increases, the alternative model predicts better; whenever it decreases, the benchmark model predicts better. The units in the graphs are not intuitive, but the time-series pattern allows diagnosis of weeks with good or bad performance. The standard error of all the observations in the graphs is based on translating MSE-T statistic into symmetric 95% confidence intervals based on the McCracken (2004) critical values. Figures indicate that newspaper information increases accuracy mainly at the beginning of the COVID pandemic. The performance is similar to the benchmark in the period before COVID and associated to high uncertainty in the performance. The SI also gives gains in March 2020, but after the initial wave, its accuracy decreases resulting in decreasing lines which become negative in the final part of the sample. Whereas the ERKs keep similar or superior performance than the benchmark after the increase associated to the beginning of the COVID, resulting in flat or increasing lines in the final part of the sample. The figures shows that the model ERK is the specification providing the largest and more stable gains.

## 4 | ECONOMIC GAINS

Finally, to shed light on the economic gains of the textual news data, we carry a simple portfolio exercise based on the naive risk parity (NRP); see Leote de Carvalho et al. (2012) and Roncalli (2013) for a discussion. The NRP uses the inverse risk approach, giving lower weight to riskier assets and larger weight to less risky assets, to construct the portfolio weights. Theoretically, NRP is based on the assumption that all the assets in the portfolio have a similar excess return per unit of risk and the weights are calculated as follows:

$$\omega_{i,t+1} = \frac{\sigma_{i,t+1}}{\Sigma_{i=1}^{N}\sigma_{i,t+1}} i = 1,\ldots,N, \tag{6}$$

where the $\sigma_{i,t+1}$ is the forecasted variance for time $t+1$ for series $i$ made at time $t$. Formula (6) is repeated for each model presented in the previous sections, in order to calculate the associated portfolio returns.

Table 4 reports the model performance accordingly to the well-known Sharpe ratio ($SR$) and the modified SR ($SR_M$) given by

$$SR_M = SR \times \left[1 + \left(\frac{Skew}{6} \times SR\right) - \frac{(Kurt-3)}{24} \times SR^2\right],$$

where $Skew$ and $Kurt$ are the portfolio returns skewness and kurtosis, respectively. The table also reports the turnover index ($TO$):

$$TO = \frac{1}{T}\sum_{T}\sum_{N}|\omega_{n,t+1} - \omega_{n,t}|, \tag{7}$$

where $w_{n,t+1}$ is the desired portfolio weight in asset $n$ at time $t+1$ and $w_{n,t}$ is the portfolio weight in asset $n$ at time $t$ before rebalancing at time $t+1$. Finally, the table reports the portfolio mean, standard deviation, skewness, and kurtosis of the portfolio.

As the table shows, the semantic importance index improves the performance of all the models that make use of them. More precisely, the models that do not use the textual information perform poorly compared to the models that make use of such information. For example,

**TABLE 4** The table reports: Sharpe ratio ($SR$), modified Sharpe ratio ($SR_M$), turnover index ($TO$), portfolio mean ($Mean$), portfolio standard deviation (Std), portfolio skewness ($Skew$), and portfolio kurtosis ($Kurt$)

| Model | *SR* | $SR_M$ | *TO* | *Mean* | *Std* | *Skew* | *Kurt* |
|---|---|---|---|---|---|---|---|
| HAR | 0.095 | 0.092 | 1.141 | 0.077 | 0.660 | −1.645 | 13.198 |
| AR | 0.091 | 0.088 | 1.133 | 0.078 | 0.741 | −1.793 | 14.706 |
| ERK | 0.116 | 0.114 | 1.146 | 0.090 | 0.652 | −1.083 | 11.589 |
| HAR-X | 0.104 | 0.101 | 1.144 | 0.083 | 0.634 | −1.377 | 12.003 |
| EW | 0.093 | 0.090 | 1.128 | 0.080 | 0.741 | −1.747 | 14.472 |
| SI | 0.094 | 0.091 | 1.124 | 0.081 | 0.745 | −1.658 | 14.166 |
| 3PFR-A | 0.115 | 0.112 | 1.140 | 0.089 | 0.659 | −1.087 | 11.501 |
| 3PFR-T | 0.106 | 0.103 | 1.139 | 0.085 | 0.647 | −1.244 | 11.586 |

*Note*: The column *Model* report the different models used to construct each portfolio: the heterogeneus autoregressive model (HAR), the autoregressive model (AR), the autoregressive extended with the semantic importance index (ERK), the equal weight combination of ARX models based on different economic-related keywords (EW), the autoregressive extended with the sentiment index (SI), the HAR model with semantic importance index (HAR-X), the three-pass regression filter with automatic proxy selection (3PFR-A), and three-pass regression filter with theory proxy selection (3PFR-T).

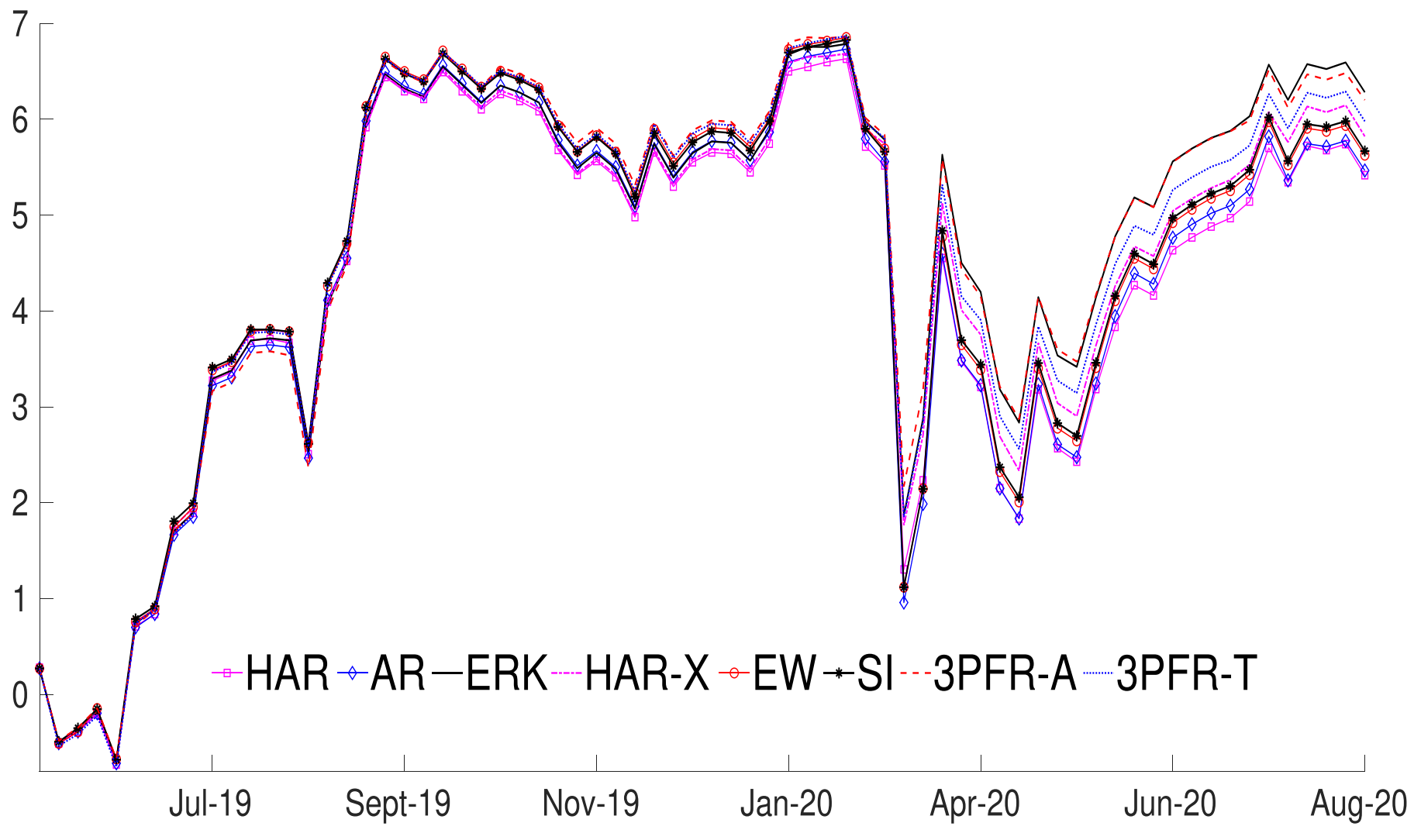

**FIGURE 7** The figure reports the cumulative sum of portfolio returns. The portfolios are constructed using the following model: the heterogeneous autoregressive model (HAR, purple squared line); the autoregressive model (AR, blue diamond line); the autoregressive extended with the semantic importance index (ERK, black continuous line); the equal weight combination of ARX models based on different economic-related keywords (EW, red circle line); the autoregressive extended with the sentiment index (SI, black circle line); the HAR model with semantic importance index (HAR-X, purple dash-dotted line); the three-pass regression filter with automatic proxy selection (3PFR-A, red dashed line); and three-pass regression filter with theory proxy selection (3PFR-T, blue dotted line)

the HAR and the AR model have lower $SR, SR_M$ with respect to the ERK and HAR-X. Moreover, the models that use the textual data have better skewness and kurtosis, decreasing uncertainty in portfolio allocation. The two 3PFRs perform better than the models without information. In this case, the automatic variable selection seems to perform better than the theoretical variable selection.

The results in Table 4 are also confirmed in Figure 7, which reports the portfolios returns cumulative sum. As the figure shows using the semantic importance index improves the cumulative returns during and after the COVID period. Regarding the EW and the SI, they do not perform well due to their inefficient use of the available information. Finally, the HAR-X gives lowest gains compared with the ERK and 3PRF-A models. This could be related to the overparameterization of the the HAR-X specification (e.g., monthly frequency).

## 5 | CONCLUSION

This paper introduces a new textual data index for predicting stock market data. The index is based on a novel methodology applied to a large set of newspaper articles to evaluate the importance of one or more general ERKs that appear in a text. The index considers three dimensions: prevalence, connectivity, and diversity. The methodology is applied to online Italian press, and 38 ERKs are selected. The resulting index is used to predict the Italian stock market and government bond returns and volatilities in the 2017–2020 period, including the inception of the COVID-19 crisis.

Our findings show that the textual index based on media news text data is able to capture the different phases and individual features of return and volatility dynamics of financial variables. Periods of large movements in the index are associated with political and economic instability. When used to predict weekly market and bond returns and volatilities, we find strong evidence of statistical and economic predictability of bond returns and volatility, as well as of stock market volatility.

## ACKNOWLEDGMENTS

The authors are grateful to Vincenzo D'Innella Capano, CEO of Telpress International B.V., and to Lamberto Celommi, for making the news data available and for the support received during the data collection process. Restrictions apply to the availability of these data, which were used under license for this study. Contact Telpress International B.V. for the permission to use them. The authors also thank participants at the EC$^2$ 2020 Conference; the 9th Italian Congress of Econometrics and Empirical Economics; Newcastle University Business School seminar series. The computing resources and the related technical support used for this study were provided by CRESCO/ENEAGRID High Performance Computing infrastructure and its staff. CRESCO/ENEAGRID

High Performance Computing infrastructure is funded by ENEA, the Italian National Agency for New Technologies, Energy and Sustainable Economic Development, and by Italian and European research programs; see http://www.cresco.enea.it/english for information. This study is also partially supported by the University of Perugia, through the program Fondo Ricerca di Base 2019, Project No. RICBA19LT ("Business Intelligence, Data Analytics e simulazioni a eventi discreti per le Smart Companies nell'era dell'Industria 4.0"). Stefano Grassi gratefully acknowledges financial support from the University of Rome 'Tor Vergata' under the "Beyond Borders" grant (CUP: E84I20000900005). Francesco Ravazzolo acknowledges financial support from Italian Ministry MIUR under the PRIN project Hi-Di NET—Econometric Analysis of High Dimensional Models with Network Structures in Macroeconomics and Finance (Grant 2017TA7TYC). Francesco Violante acknowledges financial support from the grant of the French National Research Agency (ANR), Investissements d'Avenir (LabEx Ecodec/ANR-11-LABX-0047), and Danish Council for Independent Research (1329-00011A). The funding organizations had no role in the study design, data collection and analysis, decision to publish, or preparation of the manuscript.

## DATA AVAILABILITY STATEMENT

The data that support the findings of this study are available from Telpress International B.V. Restrictions apply to the availability of these data, which were used under license for this study. Data are available from http://www.telpress.com/ with the permission of Telpress International B.V.

## ORCID

*Andrea Fronzetti Colladon* https://orcid.org/0000-0002-5348-9722

*Francesco Ravazzolo* https://orcid.org/0000-0003-0645-1788

## AUTHOR BIOGRAPHIES

**Andrea Fronzetti Colladon** Ph.D., is Associate Professor of Business Management and Analytics and Head of the Business and Collective Intelligence Lab at the University of Perugia. He is also Visiting Professor at the Department of Management of Kozminski University. Earlier, he was a Visiting Scholar at the MIT Center for Collective Intelligence. Since 2014, Prof. Fronzetti Colladon has been collaborating on several research projects at the MIT CCI, studying human communication and social interaction. In his research, Prof. Fronzetti Colladon combines methods from network science, natural language processing, and machine learning with theories from the social sciences, psychology, humanities, and linguistics to advance knowledge and discovery about management and human behavior. He is the instructor of several courses on Social Network and Big Data Analysis and the author of more than 100 scientific publications. His work has been featured in magazines and newspapers such as Harvard Business Review, Psychology

**Stefano Grassi** is an Associate Professor at the University of Rome 'Tor Vergata'. His research interests span from financial econometrics, cryptocurrency, state space models, Bayesian analysis, sequential Monte Carlo methods, DSGE models. His research outcomes are published in international peer-reviewed journals. He teaches courses in Econometrics, Financial Econometrics, Time Series Analysis, Ph.D., and Executive programs. Before joining the University of Rome 'Tor Vergata', Stefano was Lecturer at the University of Kent (UK). Stefano is CREATES International Research Fellow (Aarhus University).

**Francesco Ravazzolo** is Full Professor of Econometrics at Free University of Bozen-Bolzano and Head of the Department of Data Science and Analyticsat BI Norwegian Business School. His research focuses on commodity markets, econometrics, energy economics, financial econometrics and macroeconometrics. He has published more than 60 articles in several leading academic journals. Francesco serves the academia in several roles. He is in the editorial board of the several academic journals and he is also member of the executive committee of the Society of Nonlinear Dynamics and Econometrics and of the steering committee of the Italian Econometric Association.

**Francesco Violante** is professor at ENSAE- Institut Polytechnique de Paris. His research interests span from financial econometrics, data rich environments, derivatives pricing and risk analysis to forecasting, business analytics and environmental quantitative economics. His research outcomes are published in international peer reviewed journals. He teaches courses in Econometrics, Financial Time Series Analysis, Forecasting, Portfolio Management, Big data Strategy/Analytics in Master, PhD and Executive programs. He has been Visiting Scholar at several universities such as Aarhus University (Denmark), Singapore Management University (Singapore), ESSEC Business School (France), HEC Montreal (Canada), Université catholique de Louvain (Belgium), and London School of Economics (UK). Prior to joining ENSAE, Francesco was Professor at Sapienza University (Italy), Aarhus University (Denmark) and Maastrich University (Netherlands). Francesco is CREST research fellow (ENSAE, ENSAI and Ecole Polytechnique), CREATES International Research Fellow (Aarhus University) and Louis Bachelier Academic Fellow (Institut Louis Bachelier).

## APPENDIX A: FURTHER RESULTS

The appendix reports the forecasting results up to 4 weeks ahead. More precisely, Table A.1 reports the results for the returns, while Table A.2 reports the results for the volatility. The models are the same discussed in Section 3 and reported in Table 3.

**TABLE A.1** Forecasting results two, three, and four steps ahead for returns

| Models | FTSE MIB | BTP-2y | BTP-5y | BTP-10y | BTP-30y |
|---|---|---|---|---|---|
| Two steps ahead | | | | | |
| *RW* | *3.07* | *0.18* | *0.59* | *1.02* | *1.84* |
| AR | 0.96 | 1.00 | 1.00 | 0.99 | 0.97* |
| ERK | 1.03 | 1.01 | 1.01 | 1.00 | 0.99 |
| EW | 1.09 | 1.00 | 1.03 | 1.02 | 1.01 |
| SI | 1.09 | 1.03 | 1.04 | 1.03 | 1.00 |
| 3PFR-A | 1.08 | 1.05 | 1.03 | 1.02 | 1.01 |
| 3PFR-T | 1.40 | 1.11 | 1.09 | 1.11 | 1.04 |
| Three steps ahead | | | | | |
| *RW* | 3.01 | *0.17* | *0.57* | *1.00* | *1.86* |
| AR | 1.04 | 1.00 | 1.00 | 0.99 | 1.01 |
| ERK | 1.00 | 1.02 | 1.04 | 1.03 | 1.04 |
| EW | 1.00 | 1.03 | 1.06 | 1.04 | 1.04 |
| SI | 1.01 | 1.05 | 1.06 | 1.05 | 1.08 |
| 3PFR-A | 1.02 | 1.06 | 1.06 | 1.06 | 1.07 |
| 3PFR-T | 1.10 | 1.04 | 1.03 | 1.05 | 1.05 |
| Four steps ahead | | | | | |
| *RW* | 3.01 | *0.17* | *0.57* | *1.00* | *1.86* |
| AR | 1.00 | 1.00 | 1.00 | 1.00 | 0.99 |
| ERK | 0.98* | 0.98** | 0.98* | 1.00 | 0.99 |
| EW | 0.99 | 0.96* | 0.99 | 0.99 | 0.99 |
| SI | 1.01 | 1.02 | 1.02 | 1.01 | 1.00 |
| 3PFR-A | 1.05 | 1.02 | 1.00 | 1.00 | 1.01 |
| 3PFR-T | 0.99 | 1.10 | 1.00 | 1.00 | 1.03 |

*Note*: Relative numbers to the benchmark (RW) are given for the autoregressive extended with the semantic importance index (ERK), equal weight combination of ARX models based on different economic-related keywords (EW), autoregressive extended with the sentiment index (SI), and finally, the three-pass regression filter with automatic proxy selection (3PFR-A) and theory proxy selection (3PFR-T). The table reports the mean square prediction error (MSPE) results when forecasting Italian stock returns (FTSE MIB) and Italian bond returns (BTP-2y, BTP-5y, BTP-10y, and BTP-30y).

*The alternative model provides superior statistical forecasts at 10% significance level.

**The alternative model provides superior statistical forecasts at 5% significance level.

**TABLE A.2** Forecasting results two, three, and four steps ahead for volatility

| Models | FTSE MIB | BTP-2y | BTP-5y | BTP-10y | BTP-30y |
|---|---|---|---|---|---|
| Two steps ahead | | | | | |
| *HAR* | *0.91* | *1.09* | *0.94* | *0.83* | *0.79* |
| AR | 1.01 | 1.04 | 1.03 | 1.08 | 1.04 |
| ERK | 1.03 | 1.12 | 1.17 | 1.22 | 1.18 |
| HAR-X | 1.05 | 1.14 | 1.19 | 1.24 | 1.17 |
| EW | 1.03 | 1.03 | 1.01 | 1.01 | 1.01 |
| SI | 1.04 | 1.01 | 0.99 | 1.02 | 1.03 |
| 3PFR-A | 1.15 | 1.30 | 1.46 | 1.46 | 1.40 |
| 3PFR-T | 1.08 | 1.28 | 1.37 | 1.39 | 1.30 |
| Three steps ahead | | | | | |
| *HAR* | *0.96* | *1.16* | *0.97* | *0.88* | *0.82* |
| AR | 1.02 | 1.02 | 1.08 | 1.04 | 1.02 |
| ERK | 0.96** | 1.17 | 1.21 | 1.26 | 1.20 |
| HAR-X | 0.98** | 1.17 | 1.18 | 1.27 | 1.18 |
| EW | 0.99 | 1.04 | 1.04 | 1.02 | 1.03 |
| SI | 1.03 | 1.01 | 0.96* | 0.97 | 0.98 |
| 3PFR-A | 1.11 | 1.25 | 1.33 | 1.34 | 1.34 |
| 3PFR-T | 1.07 | 1.25 | 1.36 | 1.37 | 1.33 |
| Four steps ahead | | | | | |
| *HAR* | *0.99* | *1.12* | *1.00* | *0.88* | *0.81* |
| AR | 1.00 | 1.02 | 1.04 | 1.02 | 1.02 |
| ERK | 1.01 | 1.12 | 1.19 | 1.23 | 1.19 |
| HAR-X | 1.02 | 1.13 | 1.20 | 1.23 | 1.19 |
| EW | 1.00 | 1.06 | 1.07 | 1.03 | 1.04 |
| SI | 1.06 | 0.98* | 0.99 | 1.00 | 0.99 |
| 3PFR-A | 1.02 | 1.27 | 1.25 | 1.23 | 1.12 |
| 3PFR-T | 1.06 | 1.19 | 1.19 | 1.20 | 1.09 |

*Note*: Panel B provides MSPEs when forecasting the volatility of Italian stock and bond markets. Relative numbers to the benchmark (HAR) are given for the alternative specifications: autoregressive extended with the semantic importance index (ERK), equal weight combination of ARX models based on different economic-related keywords (EW), autoregressive extended with the sentiment index (SI), the HAR model with semantic importance index (HAR-X), and finally, the three-pass regression filter with automatic proxy selection (3PFR-A) and theory proxy selection (3PFR-T).

*The alternative model provides superior statistical forecasts at 10% significance level.

**The alternative model provides superior statistical forecasts at 5% significance level.